\documentclass[final,10pt,twocolumn]{elsarticle}

\usepackage[top=2.5cm,bottom=2.5cm,left=2.5cm,right=2.5cm]{geometry}

\usepackage{hyperref}
\usepackage{multirow}
\usepackage{amsmath,amsfonts,amssymb,amscd,amsthm,xspace}

\journal{Nuclear Instrument and Methods in Physics Research A}

\begin{document}

\begin{frontmatter}

\title{The Borexino Thermal Monitoring and Management System}

\author[myamericanaddress,myitalianaddress]{D. Bravo-Bergu\~no}

\author[RiccardoAddress]{R. Mereu}
\author[myamericanaddress]{P. Cavalcante}
\author[LNGS]{M. Carlini}
\author[Andrea]{A. Ianni}
\author[Andrea]{A. Goretti}
\author[LNGS]{F. Gabriele}
\author[myamericanaddress]{T. Wright}
\author[myamericanaddress]{Z. Yokley}
\author[myamericanaddress]{R.B. Vogelaar}
\author[Andrea]{F. Calaprice}

\address[myamericanaddress]{Physics Department, Virginia Tech, 24061 Blacksburg, VA (USA)}
\address[myitalianaddress]{INFN Sezione Milano, Via Celoria 16, 20133 Milano (Italy) - (+39) 3394636849 - \textit{david.bravo@mi.infn.it}}
\address[RiccardoAddress]{Department of Energy - Politecnico di Milano, via Lambruschini 4, 20156 - Milano, Italy}
\address[Andrea]{Physics Department, Princeton University, Princeton, NJ 08544, USA}
\address[LNGS]{INFN Laboratori Nazionali del Gran Sasso, 67010 Assergi (AQ), Italy}

\begin{abstract}
A comprehensive monitoring system for the thermal environment inside the Borexino neutrino detector was developed and installed in order to reduce uncertainties in determining temperatures throughout the detector. A complementary thermal management system limits undesirable thermal couplings between the environment and Borexino's active sections. This strategy is bringing improved radioactive background conditions to the physics signal thanks to reduced fluid mixing induced in the liquid scintillator. While thermal equilibrium has not yet been fully reached, and fine-tuning is possible, the system has proven extremely effective at stabilizing the detector's thermal conditions while offering precise insights into its mechanisms of internal thermal transport. Numerical simulations have also used this empirical data in a global detector model, providing information into present and future thermal trends.
\end{abstract}

\begin{keyword}
Neutrino detector \sep
Computational Fluid Dynamics \sep
Thermal control \sep
Radiopurity \sep
Background stability
\end{keyword}

\end{frontmatter}


\section{Introduction}
\label{sec:intro}

The Borexino liquid scintillator (LS) neutrino observatory is devoted to performing high-precision neutrino observations, and is optimized for measurements in the low energy (sub-MeV) region of the solar neutrino spectrum. Borexino has succeeded in determining all major solar neutrino flux components already with its first dataset \textit{Phase 1} (2007-10): first direct detections of \textit{pp}\cite{pp}, \textit{pep}\cite{pep}, $^7$Be\cite{7Be}, and lowest-threshold observation of $^8$B\cite{8B} at 3 MeV, as well as the best available limit in the CNO solar $\nu$ flux\cite{8B}. More recently, high-precision (down to $\sim$2.8$\%$) determinations of the aforementioned solar neutrino fluxes have been attained using new techniques and enlarged statistics from the post-LS-purification phase: \textit{Phase 2}\cite{wideband}\cite{new8B}. Geoneutrinos have also been measured with high significance (5.9$\sigma$\cite{geo}) by Borexino, thanks to the extremely clean $\overline{\nu}_e$ channel --which is expected to gain even more relevance during the \textit{Short-distance neutrino Oscillations with boreXino} (SOX) phase of the experiment. An $\overline{\nu}_e$ generator will be placed in close proximity to the detector during CeSOX, in order to probe for anomalous oscillatory behaviors and unambiguously check for experimental signatures along the phase space light sterile neutrinos might lie in\cite{SOX} \cite{Giunti}. These results are possible thanks to the unprecedented, extremely radio-pure conditions reached in the Active Volume (AV) of the detector (down to $\leq$10$^{-19}$ g($^{239}$U/Th)/g(LS)\cite{wideband}) --achieved thanks to a combination of ultra-clean construction and fluid-handling techniques, as well as dedicated scintillator purification campaigns\cite{purif}. Detailed detector response determination was made possible thanks to very successful internal calibration campaigns\cite{calib} which did not disturb the uniquely radio-pure environment. 

The unprecedented radiopurity levels reached in Borexino's LS are the key to the uniqueness of the detector's results. The conditions reached for \textit{Phase 2} after the purification campaign in 2011\cite{wideband} have raised the need for increased stability in their spatio-temporal distribution inside the detector. Indeed, mixing of the free scintillating fluid inside the IV could cause unwanted background fluctuations that, with careful management measures, may be minimized or avoided by means of external thermal environment control and stabilization. 

Section 2 of this paper will detail the correlation existent between background stability and detector thermal conditions. Section 3 will deal with the temperature monitoring solution devised and installed in Borexino in order to inform and manage the deployment of the management systems discussed in Section 4. Together, this hardware is referred to as the \textit{Borexino Thermal Monitoring and Management System} (BTMMS). Section 5 highlights the experimental results obtained by the BTMMS. Section 6 will focus on the conductive Computational Fluid Dynamics (CFD) simulations developed in order to more comprehensively understand the detector past and future thermal behavior. Finally, Section 7 will give a comprehensive view of the conclusions reached and their impact in future detector operations and physics results.

\section{Background stability}
\label{sec:stability}

\begin{figure}[ht]
\centering\small\includegraphics[width=1\linewidth]{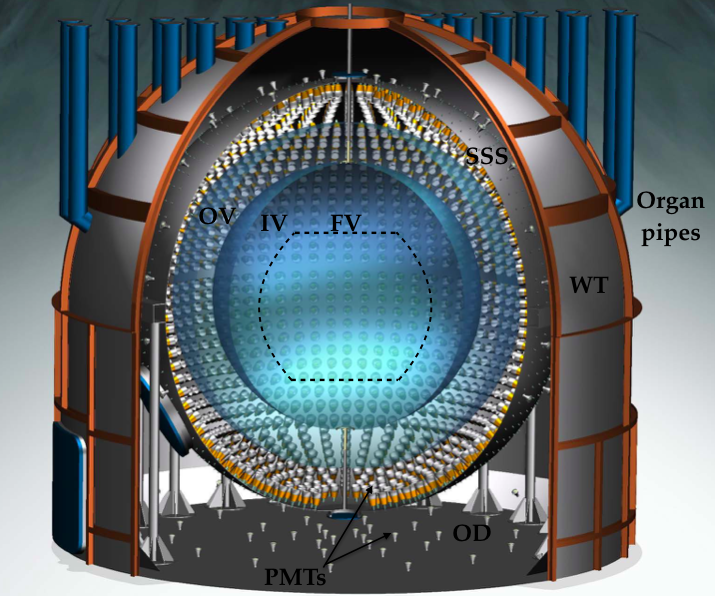}
\caption{The Borexino neutrino observatory, with its main structures annotated. See full text for details.}
\label{fig:BX}
\end{figure}

Borexino, located in the Hall C of the Gran Sasso National Laboratories' (LNGS) underground facilities (3,800 m w.e.), measures solar neutrinos via their interactions with a 278 tonnes target of organic liquid scintillator. The ultrapure liquid scintillator (pseudocumene (1,2,4-trimethylbenzene (PC)) solvent with 1.5 g/l 2,5-diphenyloxazole (PPO) scintillating solute) is contained inside a thin transparent spherical nylon Inner Vessel (IV) of 8.5 m diameter. Solar neutrinos are detected by measuring the energy and position of electrons scattered by neutrino-electron elastic interactions. The scintillator promptly converts the kinetic energy of electrons by emitting photons, which are detected and converted into electronic signals (photoelectrons (p.e.)) by 2,212 photomultipliers (PMT) mounted on a concentric 13.7 m-diameter stainless steel sphere (SSS, see Figure~\ref{fig:BX}). A software-defined, analysis-dependent Fiducial Volume (FV) is established inside the IV. The volume between the nylon vessel and the SSS is filled with 889 tonnes of ultra pure, non scintillating fluid called "buffer" acting as a radiation shield for external gamma rays and neutrons. A second, larger nylon sphere (Outer Vessel (OV), 11.5 m diameter) prevents radon and other radioactive contaminants from the PMTs and SSS from diffusing into the central sensitive volume of the detector, and segments the Inner and Outer Buffers (IB and OB). The SSS is immersed in a 2,100-tonne Water Tank (WT) acting as a \v{C}erenkov detector detecting residual cosmic $\mu^{\pm}$.

Radioactive decays within the scintillator form a background that can mimic neutrino signals. A record low scintillator contamination of $<$10$^{-18}$ g/g was achieved for $^{238}$U and $^{232}$Th. 

\begin{figure}[!htb]
\centering
\minipage{0.46\textwidth}
  \includegraphics[width=\linewidth]{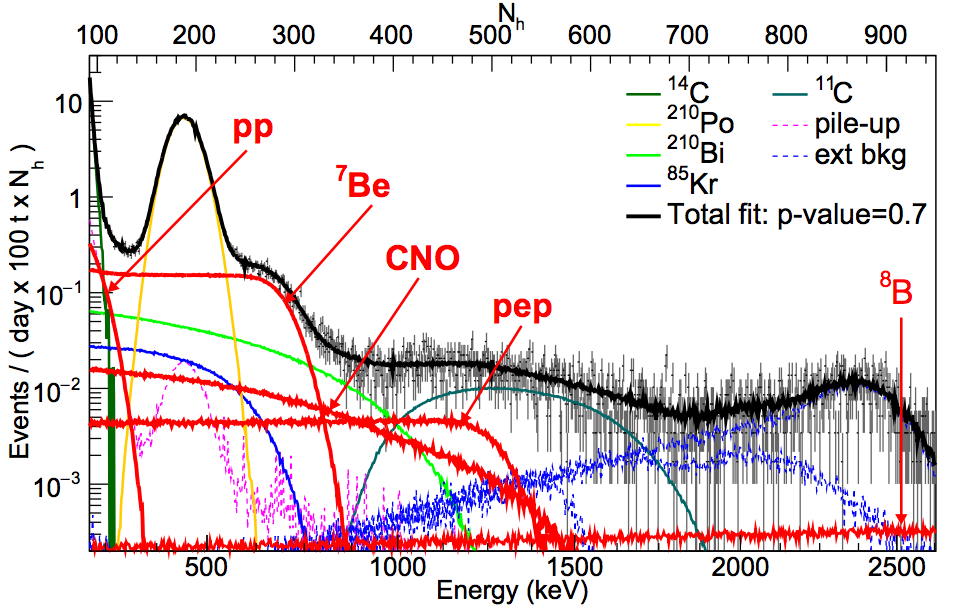}
  \caption{\small{Borexino spectrum in the main analysis energy range ($\sim$500 p.e./MeV), from \cite{wideband}. Note the near-degeneracy in the $\sim$850 keV area between $^{210}$Bi and CNO $\nu$'s spectral shapes, the only window where $\nu_{CNO}$ are prevalent over $pp$-chain $\nu$. Color version available online.}}\label{fig:spectrum}
\endminipage\hfill
\minipage{0.48\textwidth}
  \includegraphics[width=\linewidth]{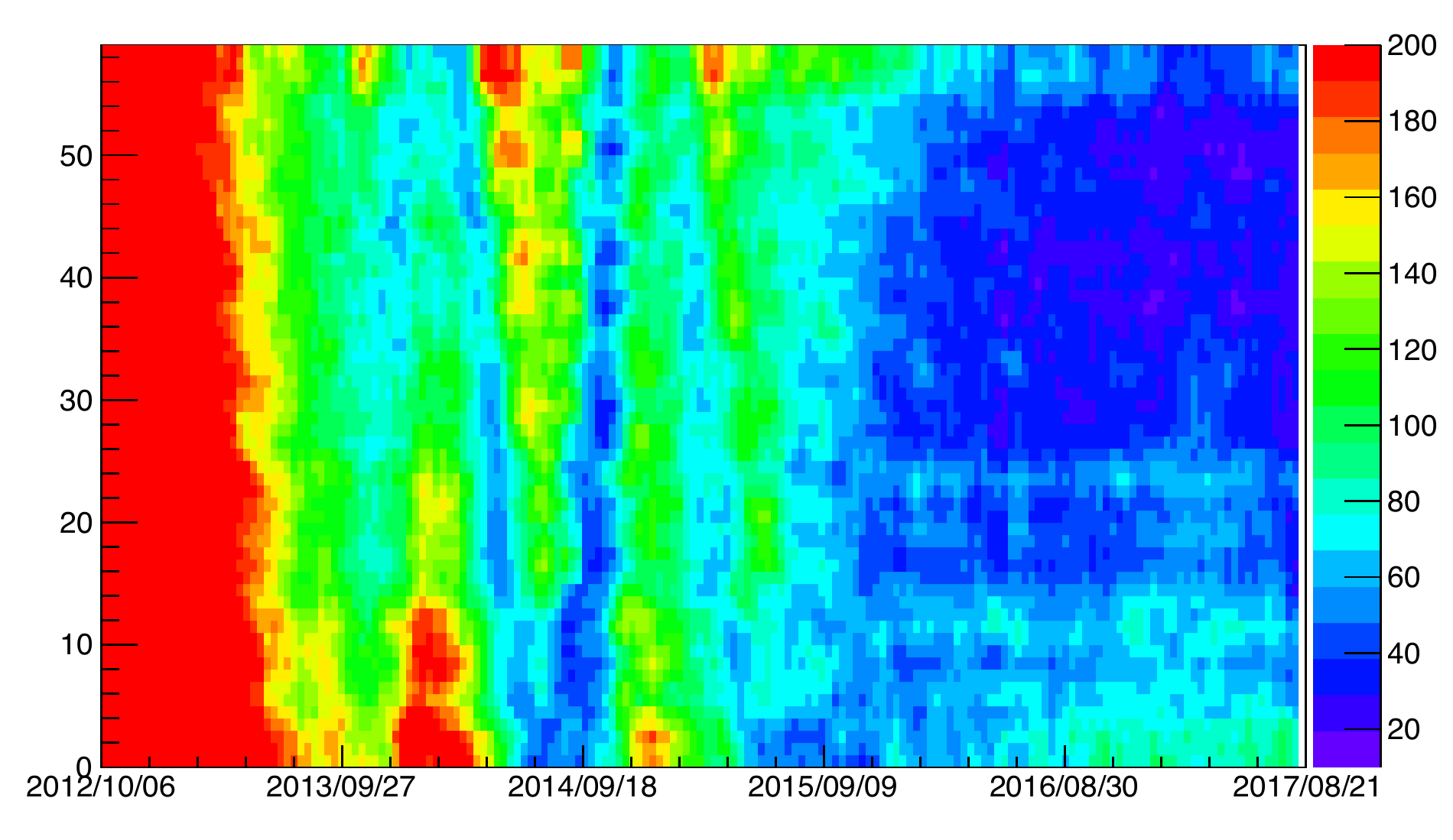}
  \caption{\small{Regional map of historical polonium concentrations (color code: cpd/100 tonnes). Expected plateau levels due to equilibrium $^{210}$Bi in the scintillator are $\sim$20 cpd/100 tonnes. The Y axis indicates regional subdivisions of a 3m-radius FV, running from its bottom to its top. Color version available online.}}\label{fig:polonium}
\endminipage
\end{figure}

Of particular importance to Borexino's future is the effort toward measuring the sub-leading, but crucial, CNO solar neutrino component ($\leq$1$\%$ of the Sun's output\cite{CNO}). Its neutrino recoil spectral shape (endpoint at $E_{max}$=1.74 MeV) places it under several intrinsic Borexino backgrounds, in particular $^{85}$Kr and $^{210}$Bi, whose spectral shapes exhibit a large degree of correlation with the solar neutrino signal --especially so for $^{210}$Bi in the $\sim$400 p.e. energy window, where the CNO rate is briefly expected to be higher than the neighboring, $^7$Be and $pep$ neutrinos (see Figure~\ref{fig:spectrum}). It is estimated a $\sim <$10$\%$ precision in the determination of the $^{210}$Bi concentration in Borexino's FV is needed, during a long enough time period, to collect the very low expected CNO $\nu$ counts ($\sim$3-5 cpd/100 tonnes).

Its decay daughter $^{210}$Po provides an accurate method for succeeding in this determination. Indeed, $^{210}$Po's $\alpha$ decay allows for it to be efficiently tagged out through Pulse-Shape Discrimination (PSD) techniques with very low inefficiencies. Conversely, the $\beta ^-$ decay of bismuth (Q=1160 keV, $t_{1/2}$=5 days) provides an indistinguishable (only statistically-subtractable) signal to $\nu_e-e^-$ elastic scatterings which cannot determine the rate down to the required uncertainty levels, due to the shape degeneracy between the bismuth and solar $\nu$ components in the so-called "bismuth valley"\footnote{The "bismuth valley" is the energy window where CNO $\nu$s are least overwhelmed by other solar neutrinos or irreducible background components, between the $^7$Be shoulder and $^{11}$C+$\nu_{pep}$, around 400 p.e..}. Indeed, once initial out-of-equilibrium $^{210}$Po levels have decayed away (initial rate $\sim$800 times higher than that of bismuth; $t_{1/2}$=138.4 days), the decay curve would asymptotically reach a plateau baseline corresponding to the secular equilibrium levels of $^{210}$Bi. This condition has been close at hand for most of Borexino's \textit{Phase 2} DAQ period --but new out-of-equilibrium, regionally-significant fluctuations in the $^{210}$Po levels have prevented reaching it (see Figure~\ref{fig:polonium}). A dedicated publication on the specialized signal analysis for $^{210}$Po-Bi and its interpretation is in preparation.

Crucially, it is known $^{210}$Pb (parent of $^{210}$Bi, $t_{1/2}$=22.3 years, off-threshold-low Q-value) exhibits higher concentrations in the IV, and consequently provides a continuous, "inexhaustible" source of $^{210}$Bi-Po. Historically, $^{210}$Po fluctuations show a correlation with large environmental temperature excursions in the experimental Hall, hinting at a possible mechanism for replenishment of out-of-equilibrium polonium in the FV: fluid mixing through temperature-driven convection from the AV's periphery around the IV toward the center. Concurrently, the regional homogeneity and stability in $^{210}$Bi concentration suggests that the underlying fluid-dynamics are slow enough to prevent most of this isotope to be transported inside the FV faster than it decays, establishing a soft upper limit in radial fluid velocity of $\frac{\mathcal{O}(m)}{\mathcal{O}(2\cdot10^6 s)}\sim <$5$\cdot$10$^{-7}$ m/s.

A yearly modulation in the asymmetry between $^{210}$Po concentration at the top and the bottom of the FV started being apparent in 2013, suggesting a correlation with external temperature trends. No azimuthal dependence is apparent, and the expected direct proportionality dependence is verified in the radial direction. Borexino's legacy thermal monitoring system, installed during the detector's construction, could provide readings on its vertical axis (8 sensors in IB and OB) and SSS surface (28 sensors), as well as selected points in the environmental air around it. They also showed the correlation between ambient temperature upsets in Hall C and background concentration spikes. However, their age and design requirements made them not the ideal system to monitor changes influencing fluid-dynamics in the IV, owing to signal coarseness, sensor position and irrecoverability for replacement or recalibration. The needed system would allow for a fine mapping of Borexino's temperature profile, as close as possible to the IV but also monitoring the thermal transport from the environment. Additonally, by determining the "ideal" temperature profile, the stabilization of the detector's thermal distribution could be attempted.

\section{Monitoring system}
\label{sec:monitoring}

\begin{figure}[ht]
\centering\small\includegraphics[width=0.7\linewidth]{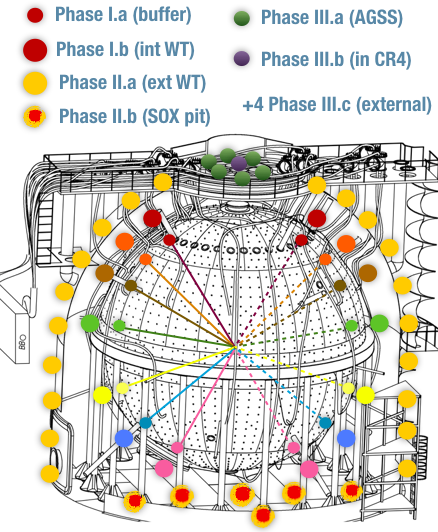}
\caption{Conceptual design of LTPS sensor positions within Borexino. Although shown to lie on the same plane, Phase I and II sensors actually show some scatter around $\phi$=0 to avoid SSS structures.}
\label{fig:positions}
\end{figure}

Borexino's natural temperature profile exhibits a stable stratification based on a temperature gradient that increases monotonically with height. Denser isotherms are present in the bottom half, indicating a sharper gradient in that region, that then smooths out toward the top, where temperatures are more uniform. Heat is exchanged with the rock, steel and concrete bottom, as well as with the air surrounding the WT, and ideally should contribute to keeping that gradient profile. In reality, local temperature inhomogeneities in the ambient air as well as seasonal upsets, deviate from this ideal situation, and generate spatial and temporal perturbations. For this reason, the Latitudinal Temperature Probe System (LTPS) is conceived as a vertical profile monitoring system, with an azimuthal resolution of 180$^{\circ}$.
\paragraph{\textbf{Layout}} The LTPS consists of three subsystems (see conceptual diagram in Figure~\ref{fig:positions}): 
\begin{enumerate}
\item \textit{Phase I}: 28 internal probes, located in the \textit{re-entrant tube} (RET) system. These ports were in principle envisioned for the insertion of small, low-activity sources for PMT and Outer Detector calibration\cite{ext_calib}, but are otherwise empty. They are located next to the PMT cable ports ("organ pipes") on top of Borexino, allowing for direct access to the SSS, up to $\sim$0.5 m into the OB.
\item \textit{Phase II}: 20 external probes located on the WT's outer wall surface, complemented by 6 probes located in a T-shaped service tunnel (1 m$^2$ section) under the detector (\textit{SOX pit}).
\item \textit{Phase III}: 6 external probes located on the WT's upper dome, one inside the calibration clean-room (CR4) located over Borexino's vertical axis, as well as several probes for exterior ambient air readings.
\end{enumerate}
The LTPS probes (see technical specifications in Table~\ref{table:Specifications}) output a voltage differential that is routed through the Signal Conditioning Box (SCB), to then be routed to the LabQuest Mini 12-bit digitizer outputting the raw data for each probe (integer number scaled to 16 bits). Once sent to the interface computer, a \texttt{C++} program converts the raw data back to voltages and temperatures according to empirically-determined calibration functions. 

\begin{figure}[ht]
\centering\small\includegraphics[width=1\linewidth]{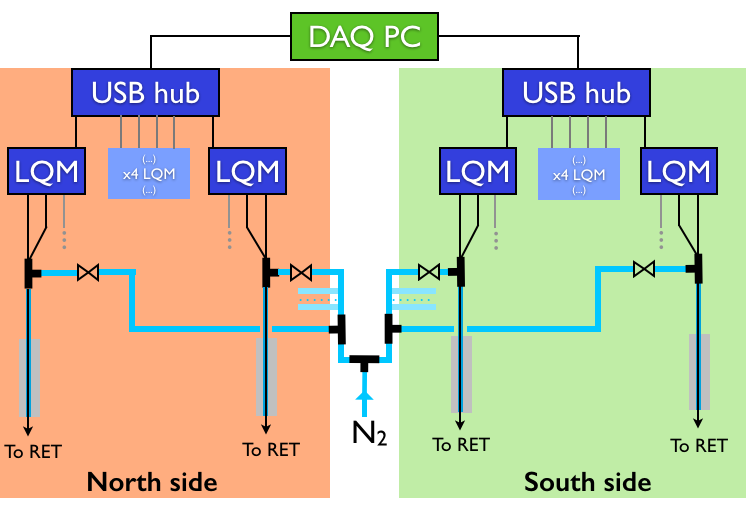}
\caption{Scheme for the 28 sensors in the LTPS Phase I layout, including electrical and gas connections. LQMs are the electronics interface supporting 3 sensors. Phase II.a shows a similar design with just one sensor per sheathing and no purging capability. Phase II.b and III have simpler routings with no N/S sides due to their special configurations. RET stands for "Re-Entrant Tubes", or the ports that reach 0.5 m inside the SSS into the OB. Color version available online.}
\label{fig:LTPS_scheme}
\end{figure}

Phase I internal probes are sheathed inside a low-friction PVC tube (10 mm OD, 8 mm ID) terminated with a small section (8 mm OD) polyethylene tube providing support for the sensor tip and avoiding disconnection between it and the sheathing. A purging and drying nitrogen flux is facilitated by slits cut at the front end of the sheathings, and fed through a manifold through their top end, in order to clear out small amounts of condensation in the ports that could damage the probes in the long run. Because of this design (see Figure~\ref{fig:LTPS_scheme}), the internal sensors are easily accessible for removal, replacement or relocation. 

The internal probes are subdivided in Phase I.a and I.b, corresponding to the probes measuring OB (0.5 m inside the SSS) and WT (0.5 m outside the SSS) temperatures, respectively. Being separated by a meter, they share the same sheathing. Phase I.a started data-taking on October 29th, 2014 and Phase I.b on April 10th, 2015.

\begin{table}[ht]
\centering
\begin{tabular}{l l}
\hline
\textbf{Specification} & \textbf{Value} \\
\hline
\textit{Order Code} & TPL-BTA \\
\textit{Temp. transducer model} & AD590JH \\
\textit{Cable length} & 30 m \\
\textit{Maximum diameter} & 7 mm\\
\textit{Range} & -50$^{\circ}$C -- 150$^{\circ}$C  \\
\textit{Specified accuracy} & $^+_-$0.2$^{\circ}$C  \\
\textit{Specified resolution} & 0.07$^{\circ}$C \\
\textit{Power} & 7.4 mA at 5 VDC \\
\textit{Response time} & 8-10 s (still water)\\
 & 45 s (stirred water) \\
 & 100 s (moving air) \\
\hline
\end{tabular}
\caption{Vernier Extra-Long Probes specifications.}
\label{table:Specifications}
\end{table}

Phase II.a and III.a sensors are located inside flexible tubing at fixed locations, for easy access as Phase I's, under the layers of the insulation system described in Section~\ref{sec:management}, and measure the external WT wall boundary condition. They went online between summer 2015 and early 2016, depending on their position. Phase II.b probes in the tunnel measure the bottom heat sink boundary condition, both at the tunnel ceiling (Borexino's bottom) as well as the absolute acquifer-controlled rock temperature on the pit floor ($\sim$6.5$^{\circ}$C) and started data-taking in autumn 2015. 

\paragraph{\textbf{Calibration}} Given the level of precision needed, a custom calibration was performed in order to improve the probes' factory specifications. This had the objectives of characterizing the probes' behavior and detect eventual individual anomalous outputs, check their short- and long-term stability, and further calibrate them in order to minimize systematic uncertainties by adding individually-tailored calibration coefficients (offset and --if necessary-- a linear term), and fine-tuning of their correction factors to an absolute reference temperature. This was achieved, in this order, with (i) characterization runs in air (low precision), (ii) absolute temperature baths (0$^{\circ}$C and ambient temperature), (iii) relative benchmarking of all probes in a single re-entrant port with constant temperature within the calibration time, for relative probe-to-probe correction factor tweaking, and (iv) absolute water bath cross-check at nominal working temperature ranges. Only the 14 Phase I.a sensors used all these techniques --lessons learned with them were extrapolated to the rest, which were only calibrated using (iv). This also avoided downtime in the already-installed Phase I.a LTPS, which would otherwise be introduced with (iii).

The calibrations yielded a $\leq$0.04$^{\circ}$C relative accuracy and a similar level of absolute precision in LTPS Phase I. Phase II and III sensors, not having been as exhaustively cross-calibrated, would exhibit a slightly worse absolute precision, but their more peripheral position renders this uncertainty negligible.
\paragraph{\textbf{DAQ Software}} Data processing in each of the 3 readout computers is based on a custom-modified version of Vernier's \texttt{NGIO\_DeviceCheck} simple acquisition code with a typical 30-minute refresh time. It is then stored in a PSQL database. Online and offline visualization and analysis tools were developed for each LPTS Phase.
\paragraph{\textbf{Technical specs results}} The LTPS sensors showed increased performance with $\sim$20$\%$ of the intrinsic jitter of the old, legacy Borexino internal thermometers ($\sim$0.01$^{\circ}$C). Long-term stability of $\leq$0.05$^{\circ}$C was also observed, in multi-hour-long periods. With these characteristics and the relative position of the Phase I.a and I.b internal sensors (separated by a distance of $\sim$1 m and measuring temperatures across the SSS' inner and outer walls) thermal transport studies to characterize the thermal transport latency could be performed. Though functional fitting of relevant transient features seen in the temporal evolution plot, the time delay between features could be quantitatively measured. An upper limit constraint of 18-24 h/m (i.e., 1-1.3 m/day) for the thermal transport speed across the SSS boundary was determined. This demonstrates a low thermal inertia to detector-wide temperature transients: extrapolating this to the bulk of the detector, even minor thermal transients would reach the IV in around 4 days.

Another relevant feature is the existence of a slight, but pervasive thermal asymmetry between the North and South sides of the detector, assumed to be mostly due to the different external environment each side is subjected to. The accompanying paper\cite{other_paper} shows the importance of these temperature asymmetries in detector fluid dynamics. These asymmetries depend on latitude, while the gradient oscillates every few months since the LTPS start-up between being larger on the North or South. From the summer of 2016, the gradient asymmetry has a pronounced asymmetry, with the North being $\sim$0.1$^{\circ}$C colder than the South. Due to environmental conditions, the North side also exhibits a "hot spot" on its lowermost Phase II.a sensor (close to the ground level of the WT wall), but this is a very localized effect with no widespread importance, and has been shown not to affect the IV in a major way.

An additional, accidental blind check of the validity of the cross-calibration performed for the Phase I probes was offered by a slight mismatch between the nominal positions of the re-entrant tubes and the actual ones: some of them were slightly displaced due to structural interference with other SSS hardware. This was most noticeable in the equatorial sensors (+7$^{\circ}$ nominal latitude), which exhibit a 3$^{\circ}$ difference, with the South sensor being lower than the North one --and therefore, with the former showing a $\sim$-0.2$^{\circ}$C chronic unexplained offset, until the positioning issue was clarified by cross-checks with the external calibration reconstructed source positions\cite{ext_calib}.

\section{Management system}
\label{sec:management}

The Thermal Insulation System (TIS) was installed in parallel to the Phase II.a probes in the May-December 2015 timeframe, with the aim of reducing the temperature transients clearly seen in data from the Phase I probes, effectively increasing the thermal resistance of Borexino's largest boundary: the WT skin. It consists of a double layer of mineral wool material (Ultimate Tech Roll 2.0 - \textit{Isover}) that covers the full surface to a depth of 20 cm. In Borexino's thermal region of interest, it has an extremely low conductivity value of $\sim$0.03-0.04 W/m$\cdot$K (see Table~\ref{table:isover}).

\begin{table}[ht]
\centering
\begin{tabular}{l c c c}
\textbf{Characteristics} & \textbf{Value} & \textbf{Units} & \textbf{EN std.} \\
\hline
Fire class & A1 & - & 13501-1 \\
Max temperature & 360 & $^{\circ}$C & 14706 \\
($>$250 Pa) & & \\
Airflow resistivity & 10 & kPa$\cdot$s/m$^2$ & 29053 \\
Acoustic absor. & 0.81 & $\alpha_w$ & ISO11654 \\
\textit{$\lambda_D$} (10$^{\circ}$C) & 0.033 & W/m$\cdot$K & 12667 \\
\textit{$\lambda_D$} (50$^{\circ}$C) & 0.040 & W/m$\cdot$K & 12667 \\
\textit{$\lambda_D$} (100$^{\circ}$C) & 0.050 & W/m$\cdot$K & 12667 \\
\end{tabular}
\caption{Technical specifications for the TIS thermal insulation material Ultimate Tech Roll 2.0 (Isover). $\lambda_D$ stands for thermal conductivity.}
\label{table:isover}
\end{table}

The exterior layer features a reflective aluminized film reinforced with an internal fiber glass grid, as well as a metallic wire mesh netting on the outside face (Ultimate Protect Wired Mat 4.0 Aluminized Isover). Metallic anchors (20-cm long) were epoxyed on the WT walls in order to support the TIS sections, with a surface density of $\sim$5/m$^2$. An approximate $>$1000 m$^2$ of detector surface were insulated, including the "organ pipes" through which the PMT cables enter the tank toward the SSS, and the interior floor of the calibration cleanroom located on the top dome. An extra $\sim$430 m$^2$ of I-beam structural elements' surface was also insulated, with just the 10-cm-thick aluminized insulation layer. Insulation started by the main surfaces, from the bottom up. Once the horizontal walls were approximately covered, I-beams followed, together with the dome's lower "rings" and largest organ pipe sections.

The Active Gradient Stabilization System (AGSS) was installed on the uppermost dome "ring" section surrounding the cusp-mounted calibration clean-room, before covering the metal wall with the insulation. It was conceptualized in order to avoid possible transient or long-term effects negatively affecting fluid stability, through the maximization of a positive thermal gradient between top and bottom of the detector, as well as the minimization of external disturbances in the top of Borexino's WT dome: the most vulnerable boundary area to environmental air temperature upsets.

The AGSS consists of twelve $\sim$18-m-long independent water loop circuits, based on 14-mm-OD copper serpentine tubing. The transfer from the circuits to the 12 input/output manifold occurs through a multilayered insulated pipe in order to guarantee a satisfactory thermal decoupling from the environmental temperature. A 3 m$^3$/h centrifugally-pumped water heater (3 kW) provides the heating power, controlled by a $\pm$0.1$^{\circ}$C-accurate controller. The system also includes an expansion tank, a mass flowmeter to manually adjust the flow and several safety items (mechanical thermostat, safety valve, pressure and flow switches). Furthermore, maximal bonding to the subjacent WT dome is ensured through copper anchors and a layer or aluminized tape, to ensure directional heat transfer toward the bottom of the heat exchanger assembly. The serpentines are horizontally distributed, in order to allow the most heated water to enter it on the upper inlet, and exit it through the lowermost outlet, ensuring maximal heat transfer at the top of the WT and keeping it constant (or reduced) toward the bottom. Nevertheless, even at constant temperature, the AGSS heating would provide stabilization, since its surface is quite small compared to the total surface of the WT dome, and it would provide a thermal anchoring effect on the topmost fluid, even if causing weak local convection in just the topmost water. 

A slow control system was implemented with data readout from 12 thermocouple precision temperature probes, in conjunction with the pressure and flow switches readouts. Constant power and constant temperature modes are available. Six of those are LTPS sensors (Phase III.a) interleaved with the serpentine in order to provide context data, while 4 outlet and 2 inlet separate sensors are used for heat transfer calculation. AGSS operation was started on January 10th, 2017, with a setpoint equal to the dome's water temperature. From late January 2017, the setpoint was raised by 0.1$^{\circ}$C/week to test the system, and by late summer 2017 it had reached a setpoint neighboring the maximum local aestival temperature.

\section{Experimental results}
\label{sec:exp_results}

\begin{figure}[ht]
\centering\small\includegraphics[width=1\linewidth]{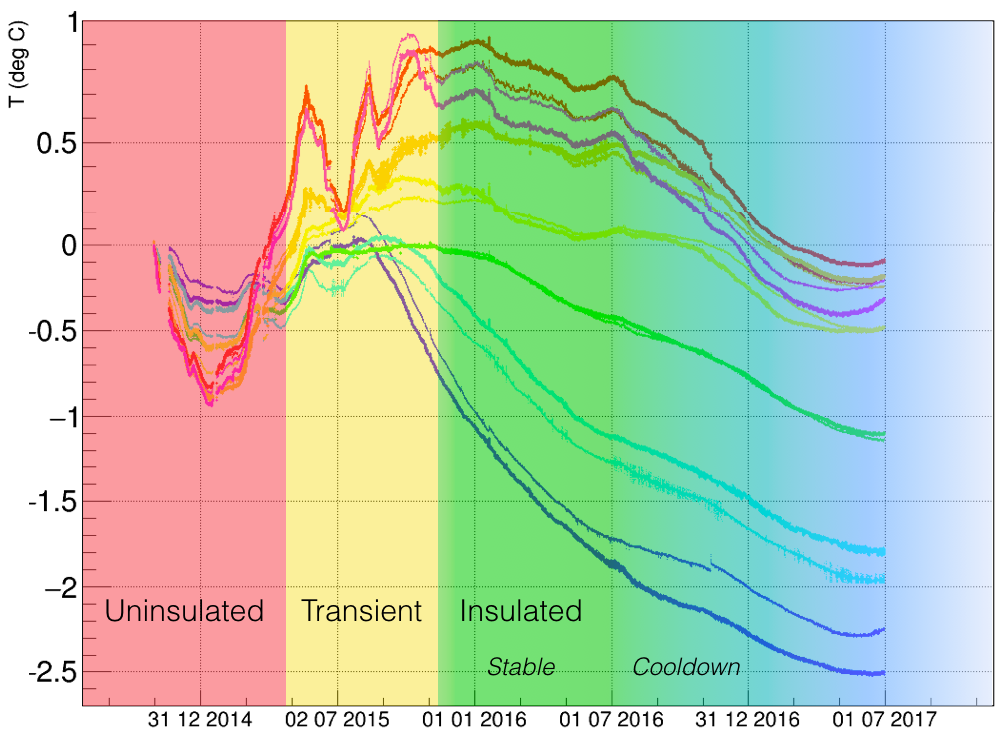}
\caption{Main periods in the October 29th, 2014 to October 21st, 2016 Phase I.a (OB) LTPS data-taking. Temperatures are offset-corrected to the same point at the start date, to then show relative drift. Color-coded curves (see online version of this work) capture the relative position of the LTPS sensors: progressively warmer colors (dark blue, cyan, green, yellow, orange, fucsia and red) indicate the increasing SSS latitudes (from $\sim$-67$^{\circ}$ to $\sim$67$^{\circ}$.}
\label{fig:periods}
\end{figure}

From the start of LTPS data taking in 2014, until mid-January 2015, a large decrease in overall temperature, and also in top-bottom gradient was noted. Historical data from the legacy thermometers show the minimum reached here ($\sim$2.2$^{\circ}$ between the extreme SSS instrumented latitudes $\pm$67$^{\circ}$) is probably the lowest ever reached since the start of Borexino data-taking in 2007. A rapid increase in gradient is then evident: while the bottom sees a slight increase of $\sim$0.15$^{\circ}$C, the top sees almost half a degree. This is considered the \textit{uninsulated} phase of the LTPS dataset. In May 2015, the TIS deployment started. This coincided with a rapid decrease in overall temperature and gradient --although at first it was mostly motivated by environmental changes in the Hall, especially in the top temperatures decrease, since TIS surface coverage was still very minor. From July 2015, seasonal temperature rise in the Hall brought the temperatures up at the same time as the loop recirculator pump for the lower half of the water in the WT was shut down. This pump increased mixing in the lower WT, and with it heat exchange, disturbing stratification. Once residual currents died down, a large decay in temperatures in the bottom half of the LTPS was evident, including in the OB. This is considered the \textit{transient} phase, until autumn (around October) of 2015, when the TIS was deployed practically globally around the detector, and the bottom cooldown was well-established. The top-bottom gradient then started an almost uninterrupted hike until July of 2016, when it reached its all-time maximum of $\sim$5.2$^{\circ}$C. TIS coverage ensured that seasonal minima in environmental air temperatures affected much less the interior of the detector. The first half-year of this \textit{fully-insulated} period showed remarkable stability in all areas of the detector, except for the foreseen (and stabilizing, in the long run) bottom cooling down. August-September of 2016 brought with it a decrease in top temperatures, which caused a break in the approximately year-long increase of the gradient. AGSS operation was started in December 2016 with the aim of partially offsetting this trend, and locking the top temperature at a constant value that should never be surpassed by environmental conditions ($\sim$17.5$^{\circ}$C). The heater temperature setpoint was gradually raised in order to have the situation fully under control and avoid de-stabilizing top heating events that could perturb the background distribution in the IV. Meanwhile, the bottom temperatures are stabilizing to the thermal sink's equilibrium temperature of ~8$^{\circ}$C, and this situation is being translated to the inside of the SSS too: the \textit{fully-insulated} period will foreseeably be subdivided into \textit{cooling} and \textit{stably-stratified} subperiods. See Figure~\ref{fig:periods} for a graphical depiction.

For illustrative purposes, a na\"ive calculation can show the cooling time constant of an ideally-insulated detector (i.e., with adiabatic walls that let no external air influence seep in, and is therefore only constrained by the heat losses through the bottom). Although simplified, this calculation represents a worst-case scenario of detector-wide, irreversible cooling, since convection will not play an important role when the lowermost fluids are stratified -- therefore, a conduction-only scenario is a very good approximation to this case, where only along-structure faster heat transport (through walls, legs...) may induce some small deviations by causing small, localized convection. However, this should only cause localized "cold finger" structures, of no or little concern.

Taking the nominal C$_{H_2O}$=4186 J/(kg$\cdot$K) and C$_{scint}$=1723 J/(kg$\cdot$K), and considering we get a mass of 280 tonnes (IV) + 1040 tonnes (OV) = 1320 tonnes of scintillator, as well as 2100 tonnes of water in the WT, we can estimate the total detector's heat capacity as:

\begin{equation}
1.32 \cdot 10^6 kg \cdot 1723 \frac{J}{kg \cdot K} = 2.27 \cdot 10^9 J/K 
\end{equation}
\begin{equation}
2.1 \cdot 10^6 kg \cdot 4186 \frac{J}{kg \cdot K} = 8.8 \cdot 10^9 J/K
\end{equation}
\begin{equation}
C_{BX}^{total}=11.1 \cdot 10^9 J/K
\end{equation}

Using the lowermost Phase II.a (and -67$^{\circ}$ I.a) sensors, which show an approximately-linear temperature drop for short enough periods ($\sim$months), and extrapolating these trends to volumes at the approximately the same temperature at each corresponding height, the worst-case heat loss through the bottom heat sink (since the data points were chosen at the beginning of the fully-insulated phase, when the inner fluids are still warm, and furthermore at the start of winter) can be estimated at $\sim<$250 W = 250 J/s $\sim$ 7.9$\cdot$10$^9$ J/year$\sim$0.3$^{\circ}$C/year. Although more realistic estimates will be shown later, this simple calculation using $\Sigma \Delta T\cdot C^{PC/H2O}_{BX}$ provides a useful upper limit.

The largest temperature gradient occurs in the bottom half of the detector, as evidenced by Phase II sensor data, changing from the approximately-constant 8$^{\circ}$C at ground level (kept stable by the aquifer located in the rock under Borexino, at $\sim$6$^{\circ}$C) to $\sim$6.5$^{\circ}$C more around the equator ($\sim$9 m higher). The stratification is, generally, much less defined in the top half, although ever present.

\section{CFD modeling}

A proper understanding of the fluidodynamic environment stemming from foreseeable thermal developments in the regions of interest inside Borexino was needed during and after the BTMMS deployment and operation. These studies are also necessary to ensure the highest internal thermal and fluid stability practicable through AGSS operations. Furthermore, development of a robust numerical study would expand to a detector-wide scale the monitoring capability that the LTPS only can provide in the discrete points where the probes make their measurements.

The commercial finite-volume solver ANSYS Fluent (v.16.2 and 17.0) Computational Fluid Dynamics (CFD) simulation package was employed in the CFDHub of the Politecnico di Milano's Computational Center. Fully convective simulations were performed as part of the study, and will be reported in the dedicated accompanying paper\cite{other_paper}. Here, the full detector simulations employing only conduction, following the Q$^3$ guidelines\cite{riccardo}, will be reported. These obeyed the desire to (i) characterize the large-scale differences the detector's configuration presented with respect to a motionless "ideal" stratification, (ii) study long-term temperature trends in the full detector under different boundary conditions, (iii) study the boundary effects of the TIS and AGSS, and (iv) identify the importance of the conducive role that structures may have as thermal bridges through the fluid. Two- and three-dimensional models using reference LTPS data were employed --even though a 2D procedure was adequate for most cases, the 3D case was an important verification for it, providing a way to study thermal transport along structures in a fully realistic geometry.

\subsection{Numerical domain and meshes}

\begin{figure}[ht]
\centering\small\includegraphics[width=1\linewidth]{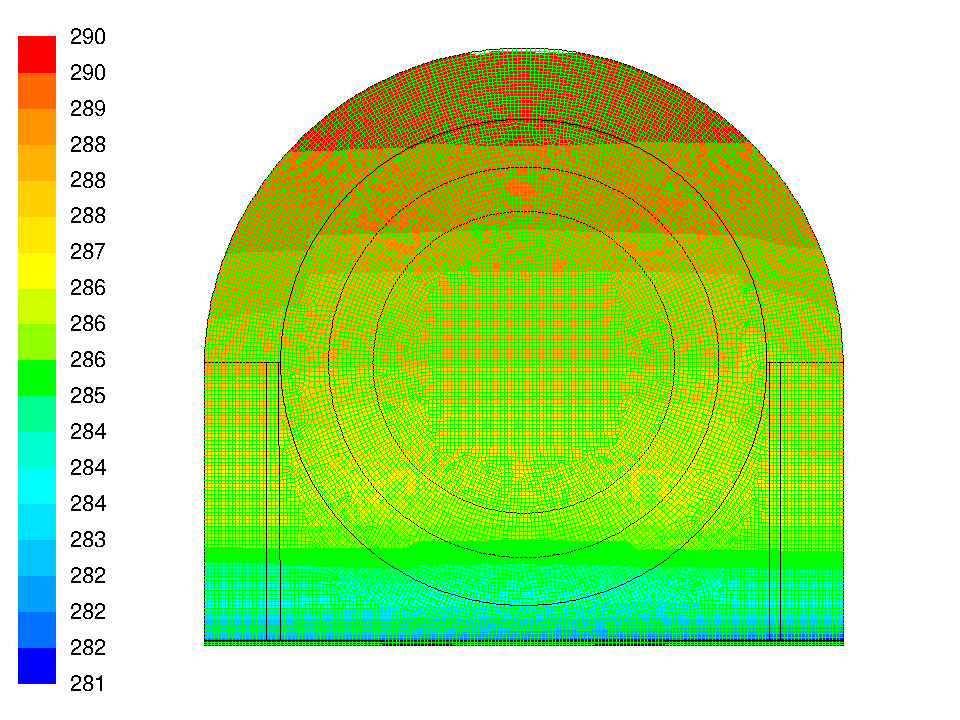}
\caption{Superimposed view of the mesh and isotherm depiction of the initialization conditions (based on LTPS data from 2015/12/15, at the end of the "\textit{Transient}" period), in K.}
\label{fig:mesh}
\end{figure}

The 2D model depicted in Figure~\ref{fig:mesh} includes the following structural elements: a Water Tank boundary (steel, 1 cm thick, 16.9 m high, 18 m OD), an SSS boundary (steel, 8 mm thick, 13.7 m OD), a North and South leg (steel, 14.3 mm thick, 32.4 cm OD, water-filled) supported on an equatorial platform (steel, 1 mm\footnote{This unrealistic thickness was chosen because of the grilled nature of the platform: even though the actual thickness is $\sim$2 cm, the porosity is estimated at $\sim$90$\%$}), Inner and Outer vessels (nylon, 125$\mu$m thick), and bottom plates (steel, 10 cm thick for the upper one on top of a smaller 4 cm thick lower one) supported on a concrete platform (14 cm maximum thickness). Water fills the WT interior and the SSS/vessels are filled with pseudocumene. An imposed temperature boundary condition mimicking the heat sink measured with the LTPS Phase II.b probes (8$^{\circ}$C) is set on the base of the model. WT wall boundary conditions vary depending on the scenario. Worst-case AGSS operation (in the sense of maximal undistributed heating) is simulated with a constant-temperature 2.1 m long band which can be turned on or off around the proper height.

The mesh is based on 10-cm side square cells on the bottom half of the water tank, transitioning to radial on the spherical dome and interior of the SSS. To avoid inaccuracies and instabilities in the region of interest in the center of the detector, a rectangular cell pattern was established again in this area, motivating irregular transition cells at around 3 m radius. In the two-dimensional model, the structural elements are just barriers with a given thermal resistance, but do not conduct along their length: this characteristic is only possible in the 3D model.

\subsection{Initial and boundary conditions}

The model was subdivided in a series of 15 different "domains", according to the height separations established by the positions of the LTPS' Phase I.a (OB), I.b (WT) and II.a (WT wall) probes. These domains were divided in \textit{North} or \textit{South} sides (and \textit{Center}, if applicable, for the volume inside the SSS). A linear interpolation was established between the known-temperature points the probes are located at. The discontinuity between domains is left to happen at roughly the position of the SSS to avoid unphysical temperature jumps happening at the most interesting regions (FV, WT periphery). The interpolation functions are:

\footnotesize
\begin{equation}
\begin{split}
T^{N/S}(x,y)=\frac{1}{A} \big[ (\Delta T(R_1) |x-|x_0|| + \Delta T(R_0) |x-|x_1||) y + \\
+ |x-|x_0|| (T_1y_1 - T_2y_0) + |x-|x_1|| (T_4y_1-T_3y_0) \big]
\end{split}
\label{eq:interp}
\end{equation}
\begin{equation}
\begin{split}
T^C(x,y)=\frac{||x|-R_0|}{A} \big[ (\Delta T^S(R_0)+\Delta T^N(R_0))y + \\
+ (T^S_4+T^N_1)y_1 - (T^S_3+T^N_2)y_0 \big]
\end{split}
\label{eq:interp2}
\end{equation}
\normalsize

where $A$ is the area between 4 interpolation points (where the temperatures $T_{1...4}$ are known from the LTPS sensors), $x_{0/1}$ and $y_{0/1}$ are the locations of each interpolation point, and $\Delta T$ is the temperature difference on each side of the interpolation square for the WT walls ($R_{0/1}$) or the SSS walls ($N/S$) at the appropriate distance from the model's vertical axis: $R_0$ refers to the Sphere's radius, $R_1$ to the WT's. These interpolation functions were properly generalized for a 3D case, in order to have a smoothly-varying temperature in all directions, through the addition of a $z$ dependence in order to account for a smooth radial function.

\subsection{Results}

\paragraph{2D conductive}

Conduction-dominated cooling was verified through the use of different boundary condition scenarios on the WT walls: ideally-insulated walls (adiabatic) and realistically-insulated walls (20-cm thick TIS) with a constant, or time-varying temperature profile (whose $\pm$2.5$^{\circ}$C modulation could be weighted or unweighted with height, since it is observed the temperature of the warmer air on the top of Hall C oscillates with larger amplitude than the thermally more stable bottom, colder air). All external air boundary conditions are based on LTPS Phase III.c and legacy sensors. A control case with no insulation was also run for a seasonally-varying gradient boundary condition. All models were run for a year of simulated time, and use the initialization thermal profile derived from the LTPS readout for December 15th, 2015.

\begin{figure}[ht]
\centering\small\includegraphics[width=1\linewidth]{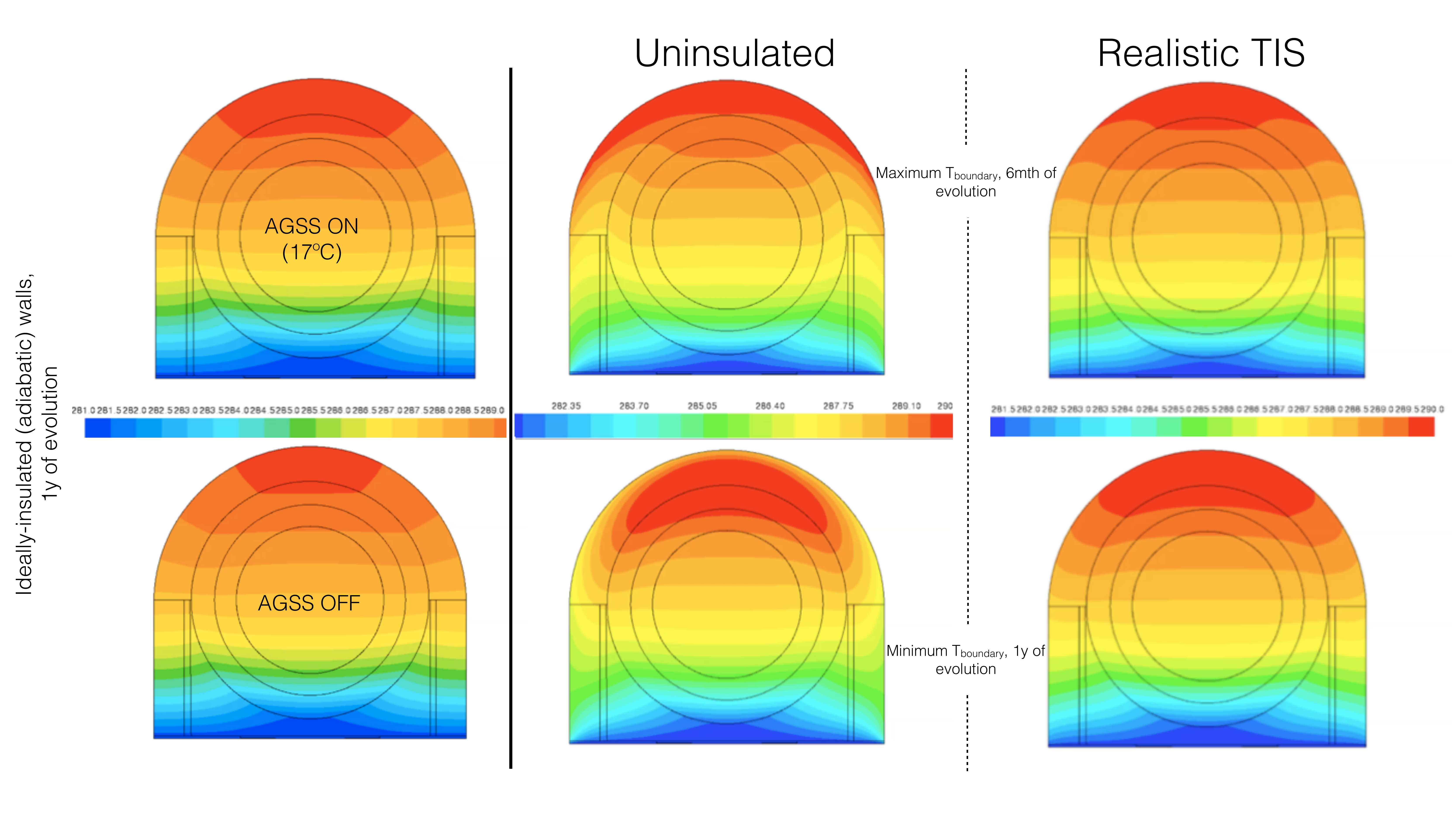}
\caption{Isotherm views of Borexino's conduction-dominated simulations. The left two figures show the stabilizing effect of the AGSS in the ideally-insulated (adiabatic) detector walls, slowing top cooling after a year of evolution. The four other figures show the difference between the height of warm (top row, 6 months of simulated time) and cold (bottom row, 1 year of simulated time) exterior boundary condition temperatures. In particular, the center two figures depict the uninsulated case with height-weighted $\pm$2.5$^{\circ}$C external oscillating boundary conditions, at their yearly maximum (top) and minimum (bottom). Identical conditions are imposed on the model on the right, which is however simulated with a TIS-like insulation layer on its exterior walls, making the seasonal changes in the periphery much milder. Scale spans 281 K to 290 K. Color version available online.}
\label{fig:evolution}
\end{figure}

Extremely similar behaviors could be seen in the center of the detector, where a "lenticular" convex feature (see Figure~\ref{fig:evolution}) grew on the cold, bottom isotherms, initially flat and horizontal, no matter what the thermal behavior was in areas closer to the boundary. This cold front progressed upwards until reaching the SSS, whereupon the isotherms regained horizontality again.

\begin{figure}[ht]
\centering\small\includegraphics[width=1\linewidth]{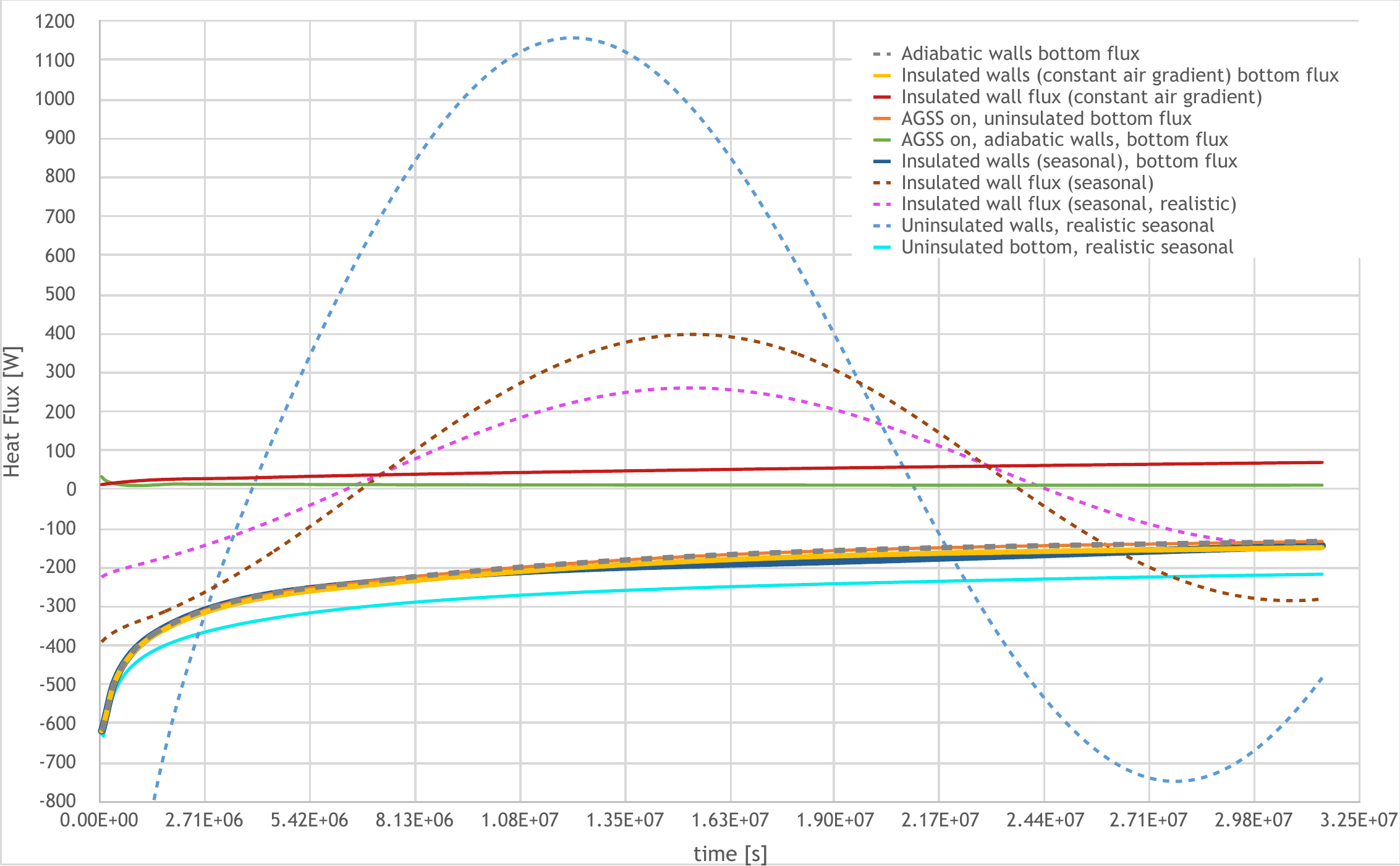}
\caption{Absolute heat fluxes through the 2D model's surfaces in different cases. Definite negative curves show heat flux through the bottom, including the perfectly-insulated/adiabatic case (discontinuous grey line), the uninsulated case (continuous light blue line) and the realistically-insulated case with constant (continuous yellow line) and seasonally-varying (continuous dark blue line) exterior temperatures. The bottom flux with adiabatic walls and activated AGSS overlaps with the one without AGSS operation at this scale (orange line under the previously indicated discontinuous grey one). Definite positive curves show instead the AGSS heat flux absorbed by the tank (continuous green line), as well as the heat exchange with a constant-temperature external environment (red). Finally, the oscillating dashed curves show heat flux through the walls in realistically-insulated cases (brown, seasonal change unweighted with height; pink, weighted), as well as the uninsulated control case (light blue, weighted seasonal change with height).  This figure's color version is available online.}
\label{fig:powerlosses}
\end{figure}

\begin{figure}[ht]
\centering\small\includegraphics[width=1\linewidth]{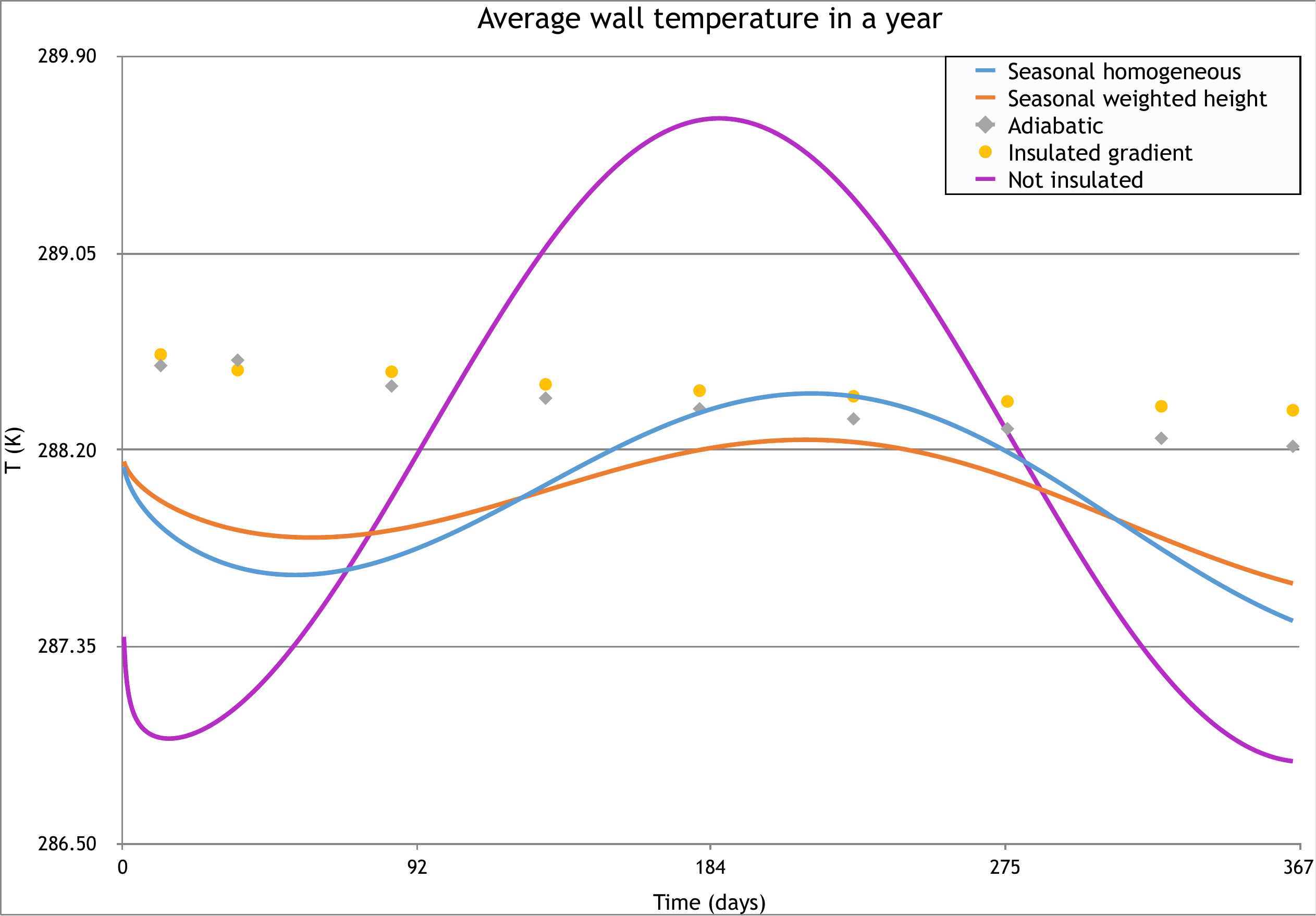}
\caption{Average wall temperatures highlighting the difference between the seasonal exterior boundary condition that changes equally at the bottom than at the top (blue) and the one weighted with height, changing maximally at the top and minimally at the bottom (orange). An uninsulated control case with weighted seasonal temperature change is also shown in purple. Dotted lines represent the scenarios with adiabatic walls not accepting any exterior heat influx or loss (grey points) and with realistically-insulated walls surrounded by a constant exterior air temperature (yellow points). This figure's color version is available online.}
\label{fig:powerlosses2}
\end{figure}

This provides a strong confirmation that, even in a worst-case situation where fluid cannot move to redistribute the heat transfer, the bottom cooling through the heat sink boundary condition retains a strongly local character, as opposed to a global effect in which the whole detector temperature drops strongly when being decoupled from the external environment by the TIS. Indeed, the only scenario where the cooling becomes global in the detector, is the one with completely adiabatic walls, as could be expected. Even in this case, the regional characteristics of the bottom cooling are retained, and only a relatively minor cooldown is seen on the top regions. As mentioned, inside the SSS, and especially in the FV, the temperature profile exhibits very minor differences between the adiabatic or realistically-insulated cases with different boundary conditions. In Figures~\ref{fig:powerlosses} and~\ref{fig:powerlosses2}, specific and integrated heat transfer time profiles are shown, identifying a maximum of $\sim$70-80 W equilibrium difference between the extreme cases of adiabatic and uninsulated walls, and the almost coincident bottom heat transfers in all insulated cases.

As mentioned in the previous section, an estimate of a maximum of $\sim$250 W of heat loss through the bottom was determined from the temperature drop in the lowermost Phase I probes during a short period of time --which allowed for linearization of the trend-- in the December 2015-January 2016 timeframe, when the detector-wide TIS installation was just finished. This estimate is well in agreement with the equilibrium heat loss through the bottom derived from these conductive models, which (after the initial $\sim$month of model stabilization from the interpolated profile) show a maximum of $\sim$-300 W, plateauing out at less than 200W of lost power in the long run.

These results yield confidence for a conduction-dominated bottom cooling that should provide a maximum temperature drop of $\sim$-0.3$^{\circ}$C/year, as estimated in Section~\ref{sec:exp_results}, with a trend that slowly reduces this drop's magnitude. This cooling is furthermore restricted to the bottom reaches of the WT, and has a negligible effect on Borexino's mid-to-upper reaches.

Differences in the time profile for the temperature inside the SSS were small in all cases ($<< \mathcal{O}$(0.1)$^{\circ}$C) and almost insignificant except for the uninsulated case. However, these conduction-only results solely provide upper limits on the thermal transport inside the Sphere, and should not be quantitatively interpreted there, as convection-dominated processes will accelerate thermal transport. Around the boundaries though, thermal behavior is much more conduction-dominated. There, temperature variations were much more prominent, where a clear "flapping" effect was seen on the isotherms (whereupon the isotherms' vertical displacement is much more pronounced close to the boundaries, and gets reduced away from them), especially on the uninsulated case, as was expected from the smoothing of outer boundary disturbances through the behavior of Poisson's equation away from those boundaries. Indeed, the interior isotherms were remarkably stable throughout seasonal variations in this conductive-only scenario, but water temperatures in the OD exhibited large oscillations that were dampened by $\sim$65$\%$ for the models with the TIS.

Stabilization of the upper gradient was also attempted by means of a simulated AGSS (two constant-temperature arcs on the location of the serpentines, simulating a maximal heating model). We imposed the approximate initialization temperature (t=0) at that height: 17$^{\circ}$C, as AGSS setpoint, and considered a worst-case cooling scenario with perfectly-insulated WT walls (adiabatic). Although the net effect in power loss through the bottom was almost negligible, highlighting the regionality of bottom cooling versus top heating, the homogenization of temperatures was less evident in the top, and kept a more ample high-temperature area around the top, effectively "holding" the top isotherms in place, although the actual heat input from the AGSS to the top water is extremely small ($\sim$0.2 W/m$^2$).

Another relevant insight is that the system, at the time of initialization, is remarkably stable conductive-wise. No major changes occur in the temperature distribution apart from the bottom cooling, a tightening of the isotherms in the lower part of the detector, and boundary effects due to the heat exchange with the environmental air, in the cases where such exchange is allowed. SSS interior temperatures remain relatively unchanged. Of course, given the symmetry of the boundary conditions, the initialization asymmetry between the North and South sides disappears quickly --nevertheless, this apparently minor feature will be shown to have great importance in the accompanying paper\cite{other_paper}. In any case, the robustness of this conductive model's results already hint at the near-stability condition the detector exhibits at the start of the fully-insulated period.

\paragraph{3D conductive}

The three-dimensional model allowed for an in-depth study of heat conduction along boundaries, which was not possible in the 2D case, along with a confirmation of the validity of the bi-dimensional model's trends. Revolution surfaces were created from the 2D mesh boundaries along the model's vertical axis, except for the legs, which were individually modeled (only 14, since it was expected they would not provide an important contribution to heat transfer). The pit under Borexino was also modeled, but its presence should not be relevant to the simulation's outcome, since only the ceiling Phase II.b sensors were considered for this setup, imposing the bottom heat sink at 8$^{\circ}$C on the WT's floor. The exterior boundary condition was kept adiabatic, although as mentioned heat conduction along the skin of the wall is possible. Two scenarios were run: with the AGSS on at 20$^{\circ}$C (the maximum foreseeable range of operations) and with the AGSS off.

Worst-case (maximal heating with homogeneous fixed-temperature band along the WT's 6th ring) AGSS operation was confirmed to remain restricted to the heated band and very narrow border areas around it, not posing any possibility of the heat creeping downward through the tank's wall (plausible in principle, given the metal's higher thermal conductivity) in a problematic fashion. A conductive-only scenario such as this should also provide a further layer of conservatism to this result, given that convection in the water should keep the heat even more localized in reality. Therefore, the AGSS system is reliably shown to offer safe operation within its intended design objectives.

Other structures that could offer a heat exchange path (legs and equatorial platform) were seen to exchange minimal heat with its surroundings, not increasing or decreasing significantly the heat conduction of the materials that surround them. This is again a worst-case scenario, since convection would only limit the heat transmission though these structures when fluids move around them. It is shown no additional heat transfer and, with it, convection, will occur because of the legs or platform, therefore making it safe to ignore them as heat-conductive elements in convective models --thereby reducing their complexity and allowing for increased efficiency. Furthermore, in the real Borexino, it is shown these elements will not cause problematic fluid-dynamic or thermal effects.

Overall thermal profiles were identical to the equivalent two-dimensional cases, which therefore are considered fully reliable for conductive-only scenarios.

\begin{figure}[ht]
\centering\small\includegraphics[width=1\linewidth]{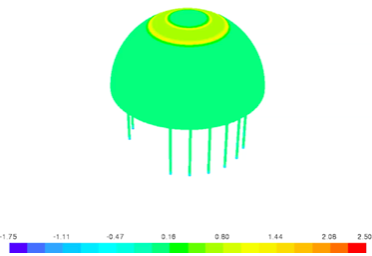}
\caption{Near-equilibrium condition of the 17$^{\circ}$C setpoint of the AGSS after a year of operation and perfect (adiabatic) insulation. Although heat exchange was greater at the beginning of the operation, because of initialization transients, non-zero heat exchange never extended beyond the limits visible in the image. Color version available online.}
\label{fig:AGSS}
\end{figure}

\section{Conclusions and prospects}

Borexino's sensitivity to solar neutrinos is determined by the rate its unprecedentedly low intrinsic background levels, which in some cases can be indistinguishable from the Compton-scattered signal.  In particular, the CNO component is very sensitive to the $^{210}$Bi levels, whose $\beta$-decay signal is in principle indistinguishable from the CNO $\nu$s scattering and overlaps with it in the spectrum's region of interest. Furthermore, it has shown temperature-correlated oscillations in its equilibrium rate with its $\alpha$-decaying daughter, $^{210}$Po. In this paper, we have discussed the deployment of the global BTMMS, composed of the precision monitoring hardware LTPS as well as the passive and active thermal management systems, TIS and AGSS, respectively. This system is aimed at increasing the detector's thermal stability with the objective of reducing scintillator mixing, exploiting the idea that background stability is directly influenced by fluid-dynamical stability. Ideally, this stabilization should reach a level which would not bring $^{210}$Po from the peripheral areas of the detector into the Fiducial Volume. The deployment of the LTPS monitoring and TIS insulating systems have unequivocally showed that (i) the increase of the top-bottom positive stratification gradient in Borexino's fluids, with relatively small North-South asymmetries, and (ii) the smoothing of the external environment's thermal upsets that transmit toward Borexino's interior, have a direct and positive impact toward the stabilization and reduction of fluid mixing. Indeed, the stratification in Borexino's interior is interpreted as the most stable ever in the detector's life, separated by one of the greatest gradients ever achieved since filling.

Conductive CFD simulations widened the level of insight into the system's stable condition by allowing for a full detector picture, anchored on the empirical data the LTPS probes provide. In particular, overcooling of the full detector has been shown not to be of concern in the foreseeable future due to the TIS installation, given the contact with the ambient air will provide enough heat exchange to stabilize the top-bottom gradient once the bottom has cooled off. In fact, the TIS is confirmed to be greatly beneficial in order to limit and delay in time the conductive heat exchange between ambient air and the WT's water, which limits the amplitude of potentially convection-inducing thermal oscillations that could propagate into the IV. The cooling process is seen to be mostly-conduction dominated. Moreover, the cooling constant obtained is in agreement with the preliminary calculations derived from first principles analysis and extrapolation of empirical LTPS data, suggesting an upper limit of lost power through the bottom heat sink. Structures have been shown not to act as significant heat bridges that could induce large-scale perturbations into an otherwise still fluid. Operation of AGSS within reasonable limits is shown to allow for the heat application to be restricted to the area of interest, and furthermore be an adequate measure to "freeze" the top isotherms in place, avoiding unwanted seasonal temperature inversions in Borexino's dome, where the stratification is the weakest.

Fully-convective, full-detector simulation is impractical even in 2D due to computing and timing limitations. Furthermore, radial segmentation in Borexino makes simulating all fluid movements in the detector unnecessary, since only the temperature boundary conditions onto the IV will impact fluid behavior therein and, with it, radioisotope transport from the periphery inwards. The accompanying paper\cite{other_paper} discusses the convective phase of the simulations, including their computational benchmarking and time-evolution models based on recorded LTPS temperatures. Additionally, an in-depth study of the implications the discovered fluidodynamic mechanisms have on background migration, as well as the data analysis techniques employed for background identification, tracking and regionalization is in preparation. Continued BTMMS operation, CFD studies to understand the fluidodynamic mechanisms at play and fine-tuned data analysis are baselining and will rule the operational and analytical strategies used in minimizing and stabilizing fluctuating radioisotope concentrations in Borexino to disentangle new precision measurements in the solar neutrino spectrum and beyond.

\section{Acknowledgements}

The Borexino program is made possible by funding from INFN (Italy), NSF (USA), BMBF, DFG, HGF and MPG
(Germany), RFBR (Grants 16-02-01026 A, 15-02-02117 A, 16-29-13014 ofi-m, 17-02-00305 A) (Russia), and NCN
Poland (Grant No. UMO-2013/10/E/ST2/00180). We acknowledge the generous hospitality and support of the Laboratori Nazionali del Gran Sasso (Italy). All numerical simulations reported in this study are performed in the HPC system of the interdepartmental laboratory 'CFDHub' of Politecnico di Milano.

\section*{References}
\footnotesize
\bibliography{BTMMS}
\normalsize
\end{document}